\documentclass[epj]{svjour}

\usepackage{graphics}
\usepackage{epsfig}
\usepackage{amssymb}
\usepackage{amsfonts}
\usepackage{amsmath}

\newcommand{\pt}{\ensuremath{p_{\mathrm{T}}} }
\newcommand{\snn}{\ensuremath{\sqrt{s_{\textrm{NN}}}}}
\newcommand{\et}{\ensuremath{E_{\mathrm{T}}} }
\newcommand{\gev}{\mathrm{GeV}}
\newcommand{\tev}{\mathrm{TeV}}
\newcommand{\mgev}{\mathrm{GeV}/c}

\begin{document}

\title{Jet Physics in ALICE}
\author{Mercedes L\'{o}pez Noriega for the ALICE Collaboration}
\institute{CERN}

\date{Received: date / Revised version: date}

\abstract{The strong suppression of high-$\pt$ hadrons observed in
heavy ion collisions at RHIC indicates the interaction of high
energy partons with a dense colored medium prior to hadronization.
We review the main results from the high-$\pt$ hadron analysis at
RHIC and what they tell us about the medium. We then concentrate on
the new possibilities that the wider kinematic range at the LHC will
offer and how they will help us to better characterize the medium
produced in these collisions. \PACS{{PACS-key}{discribing text of
that key}   \and
      {PACS-key}{discribing text of that key}}
}

\maketitle

\section{Introduction}
\label{intro}

In relativistic heavy ion collisions, large transverse momentum
partons result from the initial hard scattering of nucleon
constituents. After a hard scattering, the parton fragments to
create a high energy collimated spray of particles, that is usually
called a jet. These partons will travel through what is predicted to
be a dense colored medium, and there they are expected to lose
energy via medium induced gluon radiation~\cite{XWang94,XWang95},
this is usually called ``jet quenching". The magnitude of this
energy loss is predicted to depend strongly on the gluon density of
the medium. Therefore, measurements on how quenching changes the
structure of the jet and its fragmentation function will reveal
information about the QCD medium created in these collisions.

We start by giving a brief summary about the indications of partonic
energy loss found in the analysis of high-$\pt$ hadrons at RHIC. We
then discuss the advantages of reconstructing the full jet and how
this will be done in ALICE. We finish by describing the different
jet structure observables that we will study and what they will tell
us about the medium.

\section{High-$\pt$ Analysis at RHIC}
\label{rhic}

Clear evidences of partonic energy loss in heavy ion collisions have
been seen at RHIC. There is a strong suppression of high-$\pt$
hadrons in central AuAu collisions at $\snn = 200~\gev$ as can be
seen in the nuclear modification factor $R_{\textrm{AA}}$ for
charged hadrons and neutral
pions~\cite{JAdams03,BBack03,SAdler03,IArsene03}. $R_{\textrm{AA}}$
measures the deviation of heavy-ion collisions from a simple
incoherent superposition of pp collisions. The fact that this factor
is well below unity for $\pt > 5~\mgev$ indicates that hadron
production is suppressed by about 5 in central AuAu collisions
relative to pp collisions at that large
$\pt$~\cite{JAdams03,BBack03,SAdler03,IArsene03}. Such suppression
is however not observed in central dAu collisions at the same
collision energy indicating that the suppression observed in central
AuAu collisions is due to final-state interactions of the high
energetic partons with the dense system created in these collisions.

In the distribution of the relative azimuth $\Delta\phi$ between
pairs of charged hadrons we can see a clear dijet signal in pp and
dAu collisions as two distinct back to back Gaussian
peaks~\cite{JAdams03}. No such signal is observed in central AuAu
collisions where there is a strong suppression of the leading
fragments of the recoiling jet. Once more, the fact that this
suppression is not observed in dAu collisions indicates that the
suppression seen in AuAu collisions is due to final state
interactions of the high-energy partons with the dense colored
medium created in AuAu collisions.

Different models have incorporated energy loss via medium induced
gluon radiation in perturbative QCD calculations. This has been done
in two different ways, as multiple soft interactions
(BDMPS)~\cite{RBaier97} and as few hard scatterings
(GLV)~\cite{MGyulassy03}. In the BDMPS model the density of the
medium is characterizes by the transport coefficient $\hat{q}$.
Physically, $\hat{q}$ is the mean squared transverse momentum
generated by interactions with the medium per unit of path length
$\lambda$. In the high energy limit and for a static medium the
average radiative energy loss $\Delta E$ is proportional to
$\hat{q}$. For cold nuclear matter $\hat{q} \simeq
0.05~\gev^2/\textrm{fm}$.

In order to be able to reproduce the results from RHIC, such as the
suppression observed in $R_{\textrm{AA}}$, these models require that
the early phase of the collision has a gluon density around 30-50
times that of cold nuclear matter, which in the case of the BDMPS
model means a $\hat{q}$ between 5 and 15
$\gev^2/\textrm{fm}$~\cite{KEskola05}.

\section{Jet Reconstruction}
\label{full_jet}

As mentioned in the previous section, the effects of the medium on
the propagation of high-energy partons is studied by the analysis of
the leading particles of the jets through the $\pt$ spectra of
hadrons and through the azimuthal distribution of pairs of
high-$\pt$ particles. However, the study of the leading particles of
the jets has its limitation and they become fragile as a probe. For
extreme quenching scenarios, one observes particle emission
predominantly from the surface and therefore the sensitivity of the
leading particles to the region of highest energy density is very
limited. As a consequence of this dominance of the skin-emission,
the nuclear modification factor has very little sensitivity to the
medium properties: it is almost independent of $\pt$ and its
dependence on $\hat{q}$ becomes weaker as $\hat{q}$ increases making
the characterization of the medium based on this observable very
difficult~\cite{KEskola05}.

However, the full reconstruction of jets is, in principle, free of
such a bias and it allows the measurement of the original parton
4-momentum and the transverse and longitudinal structure of the jet.
One can then study the properties of the medium through the
modifications on the jet structure. Partonic energy loss will
manifest in a decrease on the number of particles carrying a high
fraction $z$ of the jet energy, and on the appearance of radiated
energy via an increase on the number of particles with low $z$
values. In addition, a broadening of the distribution of
jet-particle momenta perpendicular to the jet axis is expected. This
broadening is predicted to be directly related to the color density
of the medium~\cite{CSalgado04}.

\subsection{Jet Reconstruction in ALICE}
\label{reconstSection}

The ALICE detector is a general purpose heavy-ion experiment
designed to cope with the highest particle multiplicities predicted
for PbPb collisions at the LHC of $dN_{\textrm{ch}}/dy$ up to
8000~\cite{ALICE95}. The detector consists of a central part with
full azimuthal coverage in the pseudorapidity region $|\eta|<0.9$ to
measure hadrons, electrons and photons, and a forward dimuon
spectrometer to measure muons. The central region is located inside
a solenoidal magnet which allows the identification of high-$\pt$
charged hadrons with a momentum resolution better than $10\%$ up to
$100~\gev$~\cite{ALICE05}. There is a proposed electromagnetic
calorimeter that will improve jet energy resolution and will add
trigger capabilities. It will have an azimuthal coverage between $0$
and $110^{\circ}$ within $|\eta| < 0.7$~\cite{TCormier04}.

\begin{figure}
\resizebox{0.50\textwidth}{!}{
  \includegraphics{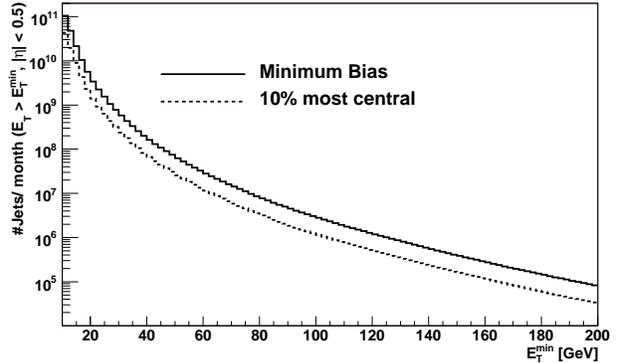}
} \caption{Predicted annual jet yield vs. energy with $|\eta|<0.5$
in PbPb collisions at $\snn = 5.5~\tev$.} \label{yield}
\end{figure}

Figure~\ref{yield} shows the predicted annual jet yields vs. energy
within the fiducial region of ALICE $\mid\eta\mid < 0.5$ for the
$10\%$ most central and minimum bias PbPb collisions for a typical
luminosity at the LHC
($5\times10^{26}~\textrm{cm}^{-2}\textrm{s}^{-1}$) for one effective
month of running ($10^{6}~\textrm{s}$). For $\et < 100~\gev$ the jet
rate $> 1~\textrm{Hz}$ is high enough to collect a sample of
$O(10^4)$ jets. For $\et > 100~\gev$ triggering will be necessary to
collect jet enriched data.

As will be shown below, jets with $\et > 50~\gev$ will allow full
reconstruction of the hadronic jets, even in the underlying
heavy-ion environment. Experimentally, jets in pp collisions are
defined as an excess of transverse energy over the background of the
underlying event within a typical cone radius $R=1$ in the
$\eta-\phi$ plane. $R = \sqrt{\Delta\eta^2 + \Delta\phi^2}$ defines
the geometrical size of the jet. In heavy-ion collisions at the LHC,
the total energy from the underlying event in a cone of $R=1$ is
expected to be of the order of 2 $\tev$, assuming $dN/d\eta = 5000$,
which is one order of magnitude higher than the highest jet energy
we want to measure. Also, this energy fluctuates by an energy which
is of the order of the jet energy. These two main limitations make
jet identification in large cones in heavy ion collisions
impossible.

However, if we take into account that about 80\% of the jet energy
is inside a cone of $R = 0.3$ and that the background energy scales
proportional to $R^2$ and its fluctuations proportional to $R$, we
can reduce the contributions from the underlying event and its
fluctuations by reducing the cone size to 0.3 or 0.4. Another way of
removing contribution of particles from the background is by
applying a transverse momentum cut to the particles inside the cone,
i.e.~by removing low-momentum particles from the cone. These cuts
may also reduced the signal but they do it to a much lesser extent.

\begin{figure}
\resizebox{0.50\textwidth}{!}{
  \includegraphics{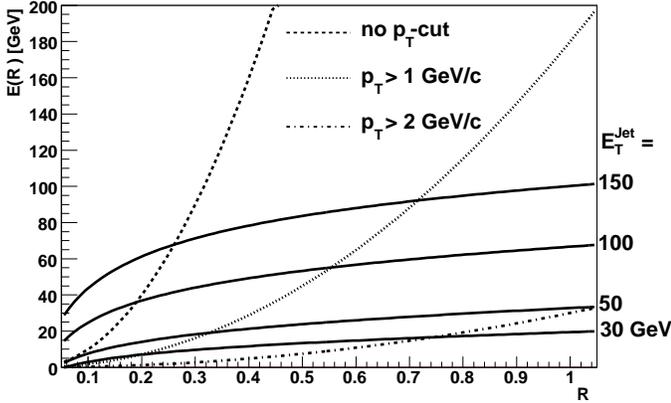}
} \caption{Background energy from charged particles within a cone of
size $R$ compared to the energy from jets of different energies for
different transverse momentum thresholds.} \label{bkgd}
\end{figure}

Figure~\ref{bkgd} compares the background energy from charged
particles within a cone of size $R$ for different $\pt$ thresholds
to the energy from jets of different energies. For $R=1$, the
background energy exceeds the jet energy even for the highest jet
energy considered in the figure. It is also clear from this figure
that the background energy can be reduced by reducing the cone size,
by applying a $\pt$ cut to the particles inside the cone, or by
doing both.

\begin{figure}
\resizebox{0.50\textwidth}{!}{
  \includegraphics{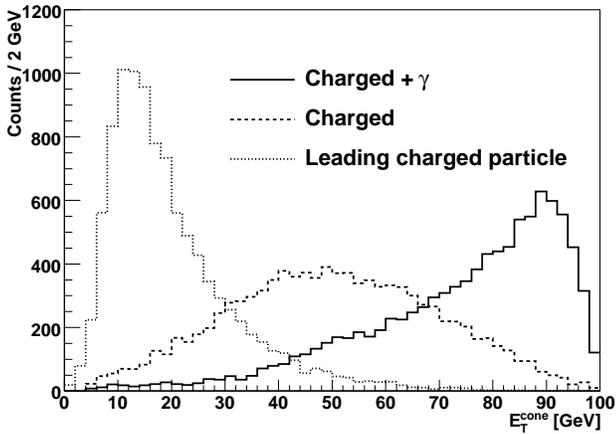}
} \caption{Spectra of reconstructed cone energy for $10~000$
generated jets with $E=100~\gev$ for different detector
configurations. A cone of size $R=0.4$ was used.} \label{spectra}
\end{figure}

However this method has its intrinsic performance limitations.
Figure~\ref{spectra} shows the spectra of reconstructed cone
energies for monoenergetic jets of $\et = 100~\gev$ generated by the
Pythia event generator~\cite{TSjost85} and reconstructed using a
cone size of $R=0.4$. The limited cone size leads to a low energy
tail even for the best case scenario in which jets are reconstructed
using charged and neutral tracks as measured by the central barrel
tracking system and the electromagnetic calorimeter. When jets are
reconstructed using only charged particles, the reconstruction is
dominated by the charged to neutral fluctuations which results in an
almost Gaussian shape with a mean value at about half the input
energy ($\Delta \et / \et \sim 50\%$).

\section{Jet Structure Observables}
\label{observables}

ALICE has studied the reconstruction of jets by simulating jet
events with the Pythia event generator and embedding them into an
underlying PbPb event generated by the Hijing event
generator~\cite{XWang91}. The generated events were passed to the
transport code GEANT3~\cite{RBrun85} which simulates the trajectory
of particles, their decays, and their interaction with the detector
materials. Signal and underlying events were then merged and passed
to the ALICE reconstruction algorithm for full reconstruction.

Two events samples were produced, one containing unquenched events
and another one containing quenched events. In the latter case,
energy loss was introduced in the Pythia simulations according to
the toy model described in~\cite{JContreras05}. An average transport
coefficient $\langle\hat{q}\rangle = 1.7~\gev^2/\textrm{fm}$ was
used. Each event sample covers 13 bins
$[\pt^{\textrm{hard}}(i),\pt^{\textrm{hard}}(i+1)]$, with
$\pt^{\textrm{hard}}(i+1)/\pt^{\textrm{hard}}(i) = 1.2$ from
$20~\gev$ to $180~\gev$. Here $\pt^{\textrm{hard}}$ is the transport
momentum of the partons in the rest frame of the hard interaction.

We used our cone algorithm to identify and reconstruct jets in these
simulated central PbPb events ($dN/d\eta = 5000$). The cone radius
was $R=0.4$ and the transverse momentum threshold
$\pt^{\textrm{min}} = 2~\gev$. After jets are reconstructed, one can
study the effects of the dense medium on the propagation of high
energy partons or their decay products by studying the modification
on the longitudinal and transverse momentum of the jet, i.e. through
the modifications to the jet structure observables. In this section
we describe different observables and present their analysis from
the generated unquenched events mentioned above. A total of 3000
events per bin of $\pt^{\textrm{hard}}$ were used. For this
jet-structure analysis, all particles inside the cone identified to
be a jet were used, even those with $\pt < 2~\gev$.

\subsection{Fragmentation Function}
\label{hbpSS}

A convenient way of representing the fragmentation function is
through the distribution of $\xi =
\ln(\et^{\textrm{jet}}/\pt^{part})$. The characteristic shape of
this distribution is known as the hump-backed
plateau~\cite{NBorghi05}. Medium induced energy loss distorts the
shape of this plateau in a characteristic way. This distortion is
caused by the decrease on the number of particles with high $z$ and
by the increase on the number of particles with low $z$.

\begin{figure}
\resizebox{0.50\textwidth}{!}{
  \includegraphics{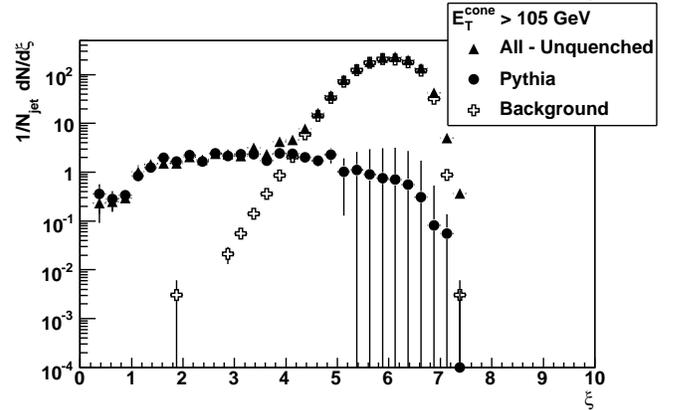}
} \caption{Reconstructed hump-backed plateau for reconstructed
energy $\et^{\textrm{cone}} > 105~\gev$. The spectra is compared to
the corresponding result for background free events and an estimate
of the background distribution.} \label{hbp}
\end{figure}

Figure~\ref{hbp} shows the reconstructed hump-backed plateau for
reconstructed energy $\et^{\textrm{cone}} > 105~\gev$ compared to
the corresponding result for background free events. Also shown in
the figure is an estimate of the background distribution. The region
$\xi<4$ corresponds to $\pt \gtrsim 1.8~\mgev$ and therefore leading
particle remnants will appear in this region where the signal to
background ratio $S/B$ is always larger than 0.1. Therefore we will
be able to measure modifications in the number of particles with
high $z$. Particles from medium-induced gluon radiation are expected
to appear predominantly in the region $4<\xi<6$. In this region
$S/B$ is of the order of $10^{-2}$ making the study of particles
with low $z$ more difficult.

\subsection{Jet Shape}
\label{shapeSection}

Jet shapes are described by the distribution of the average fraction
of energy in a subcone of radius $r$,
\begin{equation}\label{jetShape}
    \Psi(r)=\frac{1}{N_{\textrm{jet}}}\sum\frac{\pt(0,r)}{\pt(0,R)},
\end{equation}
where $R$ is the jet size.

It is clear that lowering the momenta parallel to the jet axis and
increasing the one perpendicular to it leads to an increase on the
jet size. This increase is reflected in the jet shape. Calculations
performed for a modest transport coefficient ($\hat{q}\sim
2~\textrm{GeV}^2/\textrm{fm}$) show that the energy inside a cone of
$R=0.4$ is reduced by about $5\%$ for a jet of 50 $\gev$ and by
$3\%$ for a jet of 100 $\gev$~\cite{CSalgado04}.

\begin{figure}
\resizebox{0.50\textwidth}{!}{
  \includegraphics{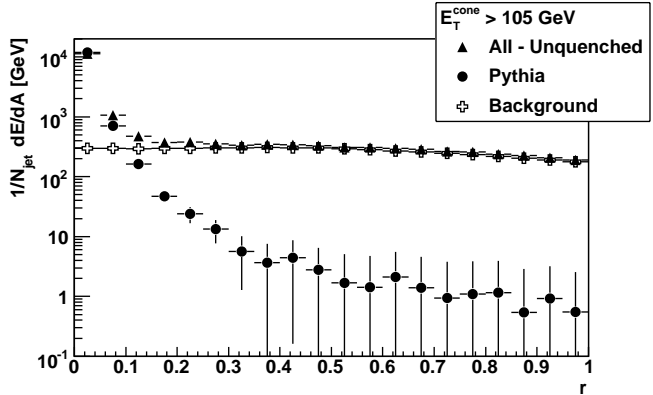}
} \caption{Reconstructed distribution of charged energy averaged
over rings $[R,R+\Delta R]$ around the jet axis normalised to the
area $A$ of the ring for reconstructed energy$\et^{\textrm{cone}} >
105~\gev$. The spectra is compared to the corresponding result from
background free events and an estimation of the background
distribution.} \label{shape}
\end{figure}

Figure~\ref{shape} shows the distribution
$(1/N_{\textrm{Jet}})dE/dR$ averaged over rings $[R,R+\Delta R]$
around the jet axis normalised to the area $A$ of the ring for
reconstructed energy$\et^{\textrm{cone}} > 105~\gev$ compared to the
corresponding result from background free events. An estimation of
the background distribution is also shown in the figure. A clear
excess of energy over the background close to the jet axis is seen.
Observing energy of the order of few $\gev$ radiated outside a cone
of $R=0.4$ will be a challenge since the signal to background ratio
decrease rapidly to $10^{-2}$ for larger radii.

\subsection{Momentum transverse to the jet axis}
\label{jtSS}

The distribution of the momentum perpendicular to the jet axis or
$j_\textrm{T}$-distribution for jets can be measured in any sub-cone
region around the jet axis. Uncorrelated particles from the
underlying event need a large angle with respect to the jet axis on
order to have a large momentum perpendicular to the jet axis.
Therefore, measuring the $j_\textrm{T}$-distribution within a small
radius preserves most of the signal and reduces significantly the
background. What one expects to measure is a broadening of this
distribution as well as an increase on its mean value due to the
extra particles produced by the radiated gluons at low
$j_\textrm{T}$~\cite{CSalgado04}.

\begin{figure}
\resizebox{0.50\textwidth}{!}{
  \includegraphics{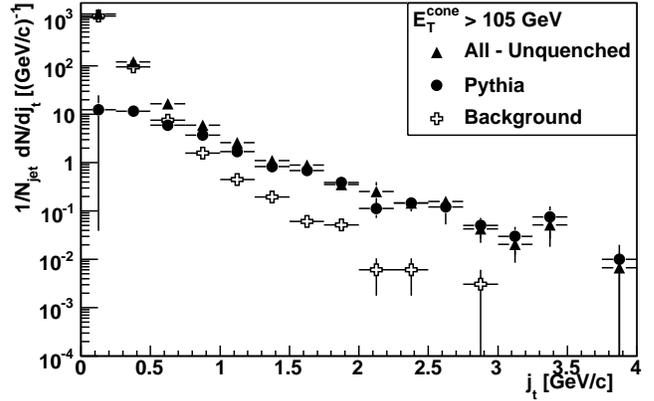}
} \caption{Reconstructed $j_\textrm{T}$-distribution for
reconstructed energy $\et^{\textrm{cone}} > 105~\gev$. Only
particles within $R<0.4$ entered the distribution. It is compared to
the corresponding result for background free events and an
estimation of the background distribution.} \label{jt}
\end{figure}

Figure~\ref{jt} shows the reconstructed $j_\textrm{T}$-distribution
for reconstructed energy $\et^{\textrm{cone}} > 105~\gev$compared to
the corresponding result for background free events. An estimation
of the background distribution is also shown in the figure. The
expected high-$j_\textrm{T}$-tail is observed.

\section{Leading Particle Correlations}
\label{RHIClike}

ALICE will also study the correlation between leading particles.
This is an important analysis for two main reasons. While the
event-by-event reconstruction of jets will be feasible for jet
energies $\et > 40~\gev$, leading hadron correlations will be in
principle possible to very low transverse momenta. Also, as shown in
section~\ref{rhic}, this kind of analysis has been performed at
lower energies, therefore we will be able to directly compare our
results with the results from RHIC that indicate a strong
suppression of the leading particles of the recoiling jet.

\section{Conclusion}

With the copious production of jets in PbPb collisions at the LHC,
ALICE will be able to identify jets using a reduced cone of
$R=0.3-0.5$ in order to reduce contributions from the background.
The properties of the created medium will then be studied through
the jet structure observables. We have shown that jet-structure
observables can actually be extracted from the sample of
reconstructed jets and that the expected characteristics are seen:
particles carrying a large fraction of the jet energy (low $\xi$),
the high-$j_\textrm{T}$-tail and the excess of energy close to the
jet axis are clearly visible in our analysis and agree with the
corresponding characteristics obtained from background-free events.

\end{document}